\documentstyle[12pt]{article}

\newcommand{\bm}[1]{\mbox{\boldmath $#1$}}
\def\be{\begin{equation}}
\def\ee{\end{equation}}

\begin{document}
\begin{flushright}
PURD-TH-95-06 \\
August 1995  \\
cond-mat/9508050
\end{flushright}
\vspace{0.4in}
\begin{center}
{\Large
Interlayer tunneling in a non-Fermi-liquid metal } \\
\vspace{0.3in}
S. Khlebnikov
\\
\vspace{0.1in}
{\it Department of Physics, Purdue University,
West Lafayette, IN 47907} \\
\vspace{0.4in}
{\bf Abstract}
\end{center}
We study the effect of interlayer tunneling in the gauge theory describing
a quasi-two-dimensional paramagnetic metal close to a second-order or weakly
first-order antiferromagnetic phase boundary. In that theory, two species of
fermions have opposite (rather than equal) charges with respect to the gauge
field.  We find that single-particle interlayer tunneling is suppressed at
low energies. The effect of pair tunneling is analyzed within the $(3-d)$
expansion. The resulting phase diagram has superconducting and
non-Fermi-liquid normal phases, and so is compatible with that of the
copper-oxide superconductors.
\newpage
Recently, a non-trivial fixed point and the associated non-Fermi-liquid
behavior have been found for a system of two-dimensional (2d) fermions
interacting with a transverse gauge field \cite{P}--\cite{CNS};
see also \cite{HNP}--\cite{V}.
The idea of non-Fermi-liquid metal has long been proposed \cite{A} to apply
to the copper-oxide superconductors. In that case, a study of interactions
between two copies (layers) of such 2d systems is important for at least two
reasons. First, the real materials are three-dimensional and one would like
to know if interlayer interactions destroy the fixed point. Second, interlayer
tunneling of pairs of electrons, in cases when single-electron tunneling is
suppressed by non-Fermi-liquid effects, was suggested as the mechanism for
high-$T_c$ superconductivity \cite{WHA,CSAS}; it seems worthwhile to explore
this possibility within the gauge theory framework.

A priori, one can imagine various types of charge assignments for fermions
with respect to the new gauge field. The correct one, for a real system,
should be determined from microscopic considerations. In the model considered
in refs.\cite{P}--\cite{CNS} all fermions have the same charge.
Here we want to consider a different model, in which there are two species of
fermions with opposite values of the charge. This charge assignment can be
derived from microphysics if the antiferromagnetic (AF) correlation length
extends over at least a few lattice spacings \cite{WWS,L}. Such situation
naturally arises in systems close to a second-order or weakly first-order
AF phase boundary. In the copper oxides, it is likely to occur at doping
concentrations lower than those for which $T_c$ is maximal---the "underdoped"
materials; however, the model itself may remain correct even at the
optimal doping concentrations. We do not have anything to add to
the derivation of this model here, but concern ourselves with its
phenomenology, namely, the phase diagram.

The main new element of our theory is that we outfit the model with the
oppositely charged fermions with local interlayer (and in-layer) interactions.
The difference in the fate of interlayer interactions between the two types of
models discussed above is indeed dramatic. In the model of
refs.\cite{P}--\cite{CNS}, pair tunneling is irrelevant at low energies,
at least in the leading order of the approximations employed.
In our model, while single-particle tunneling is suppressed
at low energies by the spin gap, pair tunneling is relevant and together
with the local in-layer interaction determines the phase diagram of the system.

Working in the leading order of the $\epsilon=3-d$ expansion \cite{CNS}, we
find two phases: superconducting and non-Fermi-liquid normal.
In the leading order, the anomalous dimension of the fermions in the normal
phase coincides with that found in refs.\cite{P,AIM}; we argue
that to hold to all orders, despite the presence of a non-vanishing repulsive
local interaction.
A phase transition with temperature between the two phases appears possible.
Thus, our model, besides being to some extent motivated from the microscopic
standpoint, has a phase diagram compatible with that of the
copper-oxide superconductors.

The new gauge field \cite{gauge} is supposed to arise as a result of
spin-charge separation, with electron dissociating into two new
quasiparticles. In the N\'{e}el phase on the square lattice we would have
on each site of sublattice A
\be
\psi_{A\sigma}=z_{\sigma} f_A^\dagger
\label{A}
\ee
and on each site of sublattice B
\be
\psi_{B\sigma}={\tilde z}_\sigma f_B^\dagger \; ,
\label{B}
\ee
where $\psi$ is the electron operator and $\sigma$ is the spin index.
The smoothly varying bosonic fields $z$ and ${\tilde z}$ represent the
staggered spin variable
\be
\mbox{\bf n} = z^\dagger \mbox{\boldmath $\sigma$} z =
-{\tilde z}^\dagger \mbox{\boldmath $\sigma$} {\tilde z}
\label{sta}
\ee
and therefore are related as
\be
{\tilde z}^{\sigma}=\epsilon^{\sigma\rho} z_\rho^*
\label{rel}
\ee
(up to an inessential phase factor) \cite{WWS,L}.
The redundancy in the number of components of $z$ is the gauge redundancy,
apparent in (\ref{A}).
Because $z$ and $z^*$ have opposite charges with respect to the gauge
field, so do $f_A$ and $f_B$.
The fermionic fields $f_A$ and $f_B$ also carry the usual electric charge,
the same amount for both species.
In the AF phase, the new gauge symmetry is spontaneously broken by a non-zero
expectation value of $z$.

We are interested, though, in the paramagnetic phase. It is more difficult
to visualize. However, the charge
assignments of the fields cannot change across the phase boundary, at least
as long as the correlation length stays sufficiently large. On both sides of
a second-order or weakly first-order phase transition a system should be
described by the same effective theory, the only difference being in the
values of the parameters.
Therefore, we describe the paramagnetic phase by the same theory containing
two oppositely charged fermions (the index A,B now merely labels the
species) and the field $z$, interacting with the gauge
field $a_{\mu}$. However, the expectation value of $z$ is now zero and all
its components are massive
("paramagnetic magnons"). The Lagrangian density describing the gauge
interactions of these fields in a single layer is given in refs.\cite{WWS,L};
our notations follow ref.\cite{L}.
By the same logic as above, even as we move further away from the phase
boundary, the interaction of the low-energy fields---the fermions and the
gauge fields---should retain its form (see Eq.(\ref{l}) below).

When there are two such layers, fields in each of them have their own
gauge transformations. The only gauge-invariant interactions between the
layers are those that correspond to tunneling of objects neutral with respect
to the gauge fields. Thus, a single fermion can only tunnel accompanied
by a quantum of the $z$ field. The corresponding lowest-dimension operators
are
\be
{\cal O}_{sA} = f_{1A}^{\dagger} f_{2A} z_2^\dagger z_1
\label{os}
\ee
and similar ones for B fermions or A$\to$B tunneling; 1,2 is the layer index.
At frequencies much lower than the spin gap (which determines the mass of $z$)
and distances much larger than the AF correlation length, the tunneling
process described by (\ref{os}) is suppressed. In that regime,
the field $z$ can be integrated out leaving only infrared-finite
renormalizations of local interactions, such as Eqs.(\ref{op}), (\ref{or})
below.

Pairs of fermions can tunnel without $z$ quanta. The pair tunneling operators
are
\be
{\cal O}_{p} = f_{1A}^{\dagger} f_{1B}^{\dagger} f_{2B} f_{2A}
\label{op}
\ee
and its hermitean conjugate.
Within the effective low-energy theory, these are viewed as local operators in
the 2d space and time, analogous to the reduced BCS Hamiltonian.
(For example, in the effective theory
there is no objection to taking two holes at the same space point, as
in (\ref{op}).)
The pair tunneling leads to a renormalization of the in-layer four-fermion
interactions
\be
{\cal O}_{r1} = f_{1A}^{\dagger} f_{1B}^{\dagger} f_{1B} f_{1A} \; ,
{}~~~~~~~{\cal O}_{r2} = f_{2A}^{\dagger} f_{2B}^{\dagger} f_{2B}f_{2A}\; ,
\label{or}
\ee
which we then have to include in our analysis.
The particle-hole tunneling operators not involving $z$ fields are
\be
f_{1A}^{\dagger} f_{1A} f_{2A}^{\dagger} f_{2A},
{}~~~~~f_{1A}^{\dagger} f_{1A} f_{2B}^{\dagger} f_{2B}
{}~~~~~~~\mbox{etc.}
\label{ph}
\ee
but these will turn out to be irrelevant at small values of
the corresponding coupling constants.

The physical situation we want to describe is as follows.
We assume that the gauge field has become deconfining due to effects of
the gapless fermions \cite{dec}. Because in our
model A and B fermions have opposite gauge charges, A and B particles moving
with opposite momenta make parallel currents and are attracted to each other.
This attraction increases their chances to be near each other and thus
effectively enhances the local interactions (\ref{op}), (\ref{or}).
We then have to understand if this enhancement can be limited by a repulsive
effect of the local interactions themselves.

There may be different ways to describe the physics outlined above. Here
we use the perturbation theory built with the resummed gauge propagator
\cite{HNP}, incorporating the Landau damping, but with the usual free
fermion propagator (in contrast to the $1/N$ method of refs.\cite{P,AIM}).
The advantage of this approach is that the strongest infrared effects of
the local four-fermion interactions are easy to isolate. These are the
infrared-singular renormalizations coming from the BCS channel.
Another useful device is the $(3-d)$ expansion \cite{CNS}.
At $d\to 3$, the infrared divergences due to the gauge interaction are also
logarithmic, and the logarithms can be conveniently handled by the
renormalization group.
Rather than using a modified fermion propagator, the
$(3-d)$ expansion introduces an additional renormalizable parameter---the
Fermi velocity. Such a trade-off is familiar from the $\lambda\phi^4$
scalar theory: in the $(4-d)$ expansion $\lambda$ is a running
coupling, while in the large-$N$ expansion directly in $d=3$ it is not.

We thus consider the Lagrangian
\begin{eqnarray}
L & = & \sum_{l=1,2} \left[
-\frac{1}{4\tilde{e}^2}
(\partial_{\mu} a_{l\nu} - \partial_{\nu} a_{l\mu})^2
+F_l^{\dagger} \left(
                    i\partial_t + \sigma_3 a_{l0}
                    +\frac{1}{2m} (\nabla-i\sigma_3 \bm{a}_l)^2 + \zeta
             \right) F_l
                     \right] \nonumber \\
  &  & \mbox{}
-g({\cal O}_{r1} + {\cal O}_{r2}) - g'({\cal O}_p + {\cal O}_p^{\dagger})\; ,
\label{l}
\end{eqnarray}
where $l$ is the layer index, $\tilde{e}^2$ is the gauge coupling
(which is not renormalized by the gauge interaction \cite{GW}),
$F_l=(f_{lA}, f_{lB})^T$, $\sigma_3$ is the third Pauli matrix, and $\zeta$
is the chemical potential. We follow
the evolution of three couplings: $g$, $g'$ and
$\alpha\equiv \tilde{e}^2v_F/4\pi$ where $v_F$ is the Fermi velocity.
The one-loop renormalization of these couplings in $d\to 3$ dimensions is
\begin{eqnarray}
g(\mu) & = & g_b+ \frac{6 g_b\alpha_b}{\epsilon^2 \pi \mu^{\epsilon/3}}
 - c_b (g_b^2+{g'_b}^2) \ln\frac{\Lambda}{\mu} \label{g} \\
g'(\mu) & = & g'_b +  \frac{6 g'_b\alpha_b}{\epsilon^2 \pi \mu^{\epsilon/3}}
- 2 c_b g_b g'_b \ln\frac{\Lambda}{\mu} \label{g'} \\
\alpha(\mu) & = & \alpha_b  - \frac{\alpha_b^2}{\epsilon\pi
\mu^{\epsilon/3}}  \label{alpha}
\end{eqnarray}
where we have used the sharp infrared cutoff at frequency $\mu$;
the subscript $b$ denotes the bare parameters; $\epsilon=3-d$,
$c_b=p_F^2/2\pi^2v_{Fb}$, and we do not write terms suppressed by powers
of the ultraviolet cutoff $\Lambda$.
The second terms on the right-hand sides of (\ref{g}), (\ref{g'}) come
from the gauge interaction in the BCS channel, while the last terms from
the BCS diagrams with the local four-fermion vertices.
To reiterate the important point, because in our
model $f_A$ and $f_B$ have opposite gauge charges, the renormalization of
$g$, $g'$ due to the gauge interaction is of the opposite sign
compared to the model of refs.\cite{P}--\cite{CNS} and makes small $g$, $g'$
grow in the infrared.
Eqs.(\ref{g}), (\ref{g'}) do not contain corrections due to the renormalization
of the fields $f$. These are of the order $\epsilon$ compared to the second
terms on the right-hand sides of (\ref{g}), (\ref{g'}) and therefore are
subleading in the $(3-d)$ expansion.

We have found that the one-loop correction to four-fermion interactions from
the gauge interaction in the $2p_F$ channel does not contain an infrared
divergence in our perturbation theory at $d\to 3$. Due to the renormalization
of the fields, the particle-hole operators (\ref{ph})
are then irrelevant as long as the bare
values of couplings with which they are added to the Lagrangian (\ref{l})
are sufficiently small.

Eqs.(\ref{g})--(\ref{alpha}) also show that the naive
infrared dimensions of $g$, $g'$ are zero, while that of $\alpha$ is
$\epsilon/3$ (not $\epsilon$, its mass dimension, as stated in
ref.\cite{CNS}). So, we introduce
\be
\lambda(\mu) = \frac{\alpha(\mu)}{\mu^{\epsilon/3}} \; .
\label{lambda}
\ee
It is also convenient to introduce $u=cg$ and $u'=cg'$ instead of $g$ and
$g'$. Notice that $c$ contains renormalizable quantity $v_F$
but the difference between $c(\mu)$ and $c_b$ is inessential to the leading
order of the $\epsilon$ expansion. The one-loop beta-functions are
\begin{eqnarray}
\frac{du}{d\ln\mu} & = & -\frac{2u\lambda}{\epsilon\pi} + u^2+{u'}^2
\label{bg} \\
\frac{du'}{d\ln\mu} & = & -\frac{2u'\lambda}{\epsilon\pi} +2uu'
\label{bg'} \\
\frac{d\lambda}{d\ln\mu} & = & - \frac{\epsilon}{3} \lambda +
\frac{\lambda^2}{3\pi} \label{bl} \; .
\end{eqnarray}
Note that the first two of these are singular at $d\to 3$ (renormalization
directly at $d=3$ would produce $\ln^2\mu$).

In the infrared, $\lambda$ runs to the stable fixed point \cite{CNS},
$\lambda=\epsilon\pi$, which describes the same
non-Fermi-liquid behavior as the modified fermion propagator of
refs.\cite{P,AIM} and, probably, the fixed point of ref.\cite{NW}.
Substituting the fixed point value of $\lambda$ into
Eqs.(\ref{bg}),(\ref{bg'}), we obtain the equations for infrared
flows, $t=-\ln\mu$,
\begin{eqnarray}
du/dt & = & 2u - u^2-{u'}^2 \label{fg} \\
du'/dt & = & 2u' - 2 uu' \label{fg'} \; .
\end{eqnarray}
The resulting flows (which can be found analytically) are plotted in
Fig.1. The flows are symmetric under $u'\to-u'$,
so we assume $u'$ non-negative. There is a critical line at $u=u'$.
If $u_b>u'_b$, the couplings run to the infrared stable fixed point
$u=2$, $u'=0$. This represents a non-Fermi-liquid normal state with
repulsive in-layer four-fermion interaction.
The interlayer electric charge transport in this state, either by single or
pair tunneling, is suppressed.
If $u_b<u'_b$, the couplings run without limit, so we expect that the system
generates a mass gap. Because the couplings run to negative values of $u$
(attraction), we expect that this phase is superconducting. It should be
similar in properties to that of ref.\cite{CSAS}.
Because the mass gap cuts off the infrared renormalization, the
single-particle and the particle-hole tunneling operators are partially
recovered in this state.

The microscopic mechanism of superconductivity may be purely electronic
(in which case $u_b$ is
the Coulomb pseudopotential usually known as $\mu$), or a value of
$u_b$ smaller than $u'_b$ may be achieved by a partial compensation of the
Coulomb repulsion by some attractive effect, due to phonons, spin
fluctuations etc.

By analogy with the BCS theory, we expect that the critical line (in the
BCS case, point) moves with temperature so that more points in the $(u,u')$
plane fall in the domain of attraction of the normal phase. For a given point
$(u_b,u'_b)$, the temperature at which the line passes through that point is
the critical temperature of the superconducting to normal transition.

Because of the singularity at $\epsilon\to 0$ in the beta-functions
(\ref{bg}), (\ref{bg'}),
the first, linear in $u$, $u'$ terms in (\ref{fg}), (\ref{fg'}) are not
proportional to $\epsilon$,
and the fixed point value of $u$ is not suppressed by $\epsilon$. This
does not mean a breakdown of the $\epsilon$ expansion. All the terms on
the right-hand sides of Eqs.(\ref{bg}), (\ref{bg'}) arise from the exceptional,
BCS ladder diagrams, which only contribute to the beta-functions at one loop
(cf. a similar discussion of the $1/N$ expansion in
ref.\cite{P}.) In particular, the usual argument from the Fermi-liquid theory
shows that there will be no terms of higher order in $u$, $u'$ not
accompanied by powers of $\tilde{e}^2$ in any of the three beta-functions.
The renormalization of $\lambda$ is restricted even further.
The renormalization of $v_F$ is determined from the fermion self-energy.
Because in the $(3-d)$ expansion the gauge interaction adds only logarithmic
singularities, it dressing vertices and internal lines of infrared-convergent
Fermi-liquid graphs cannot turn them into divergent ones. Corrections to the
gauge vertex are suppressed at low energies in $d>2$ \cite{AIM}. Then, the
only self-energy graphs that produce an infinite renormalization of $v_F$ are
those in which a gauge line encloses the rest of the graph.
These are summed up by the one-loop beta-function (\ref{bl});
the finite renormalizations replace $v_{Fb}$ with some $v_F^*$ in the bare
coupling $\lambda_b$.
By appealing again to the Fermi-liquid theory, we argue that fermions with
local interactions can be integrated
out with essentially the same effect as those without it \cite{GW}, so that
the gauge coupling $\tilde{e}^2$ is still not renormalized.
Hence, we argue that the beta-function (\ref{bl}) for $\lambda$ is in fact
exact. The fact that the fixed point value of $u$ does not vanish at $d\to 3$
suggests the existence of a non-Fermi-liquid normal state in more isotropic
three-dimensional metals close to an AF phase boundary, in which the
quasiparticle weight vanishes logarithmically at small energies but the
four-fermion repulsion remains constant.

In summary, we have studied the effect of interlayer tunneling in the
2d gauge theory with two oppositely charged species of fermions,
arising as a low-energy energy description of quasi-2d paramagnetic metals
close to an AF phase boundary.
In the leading order of the $(3-d)$ expansion, we have
found two phases, superconducting and non-Fermi-liquid normal, suggesting
that our theory may apply to the copper-oxide superconductors.

The author is grateful to S. Chakravarty for many valuable comments and
to S. Love for discussions during the course of the work.
This work was supported in part by the U.S. Department of Energy under
grant DE-FG02-91ER40681 (Task B), by the National Science Foundation under
grant PHY 95-01458, and by the Alfred P. Sloan Foundation.

\newpage
\begin{center} FIGURE CAPTION \end{center}
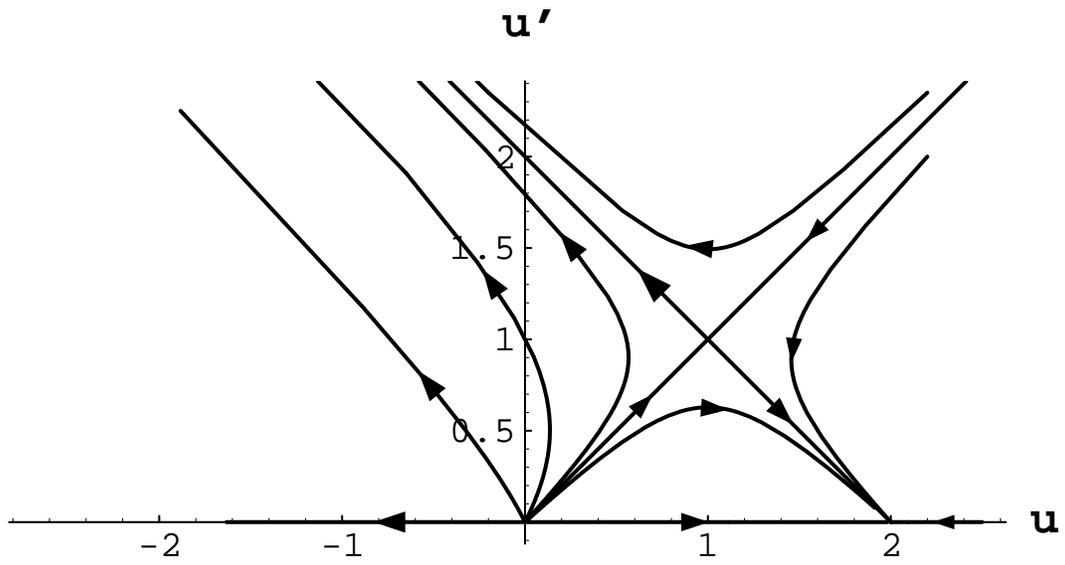
FIG. 1. Renormalization-group flows at the fixed point value of $\lambda$
for the model in text. The arrows point towards the infrared.


\begin{thebibliography}{99}
\bibitem{P} J. Polchinski, Nucl. Phys. B 422, 617 (1994).
\bibitem{NW} C. Nayak and F. Wilczek, Nucl. Phys. B 417, 359 (1994).
\bibitem{AIM} B. L. Altshuler, L. B. Ioffe, and A. J. Millis, Phys. Rev. B
50, 14048 (1994).
\bibitem{CNS} S. Chakravarty, R. E. Norton, and O. V. Sylju{\aa}sen, Phys.
Rev. Lett. 74, 1423 (1995).
\bibitem{HNP} T. Holstein, R. E. Norton, and P. Pincus, Phys. Rev. B 8,
2649 (1973).
\bibitem{R} M. Yu. Reizer, Phys. Rev. B 39, 1602 (1989).
\bibitem{L} P. A. Lee, Phys. Rev. Lett. 63, 680 (1989).
\bibitem{HLR} B. I. Halperin, P. A. Lee, and N. Read, Phys. Rev. B 47, 7312
(1993).
\bibitem{GW} J. Gan and E. Wong, Phys. Rev. Lett. 71, 4226 (1993).
\bibitem{V} C. M. Varma, Phys. Rev. Lett. 75, 898 (1995).
\bibitem{A} P. W. Anderson, in {\em Frontiers and Borderlines in
Many-Particle Physics}, edited by R. A. Broglia and J. R. Schrieffer
(North Holland, Amsterdam, 1988); Science 256, 1526 (1992);
C. M. Varma, P. B. Littlewood,
S. Schmitt-Rink, E. Abrahams, and A. E. Ruckenstein, Phys. Rev. Lett.
63, 1996 (1989); {\it ibid.} 64, 497(E) (1990).
\bibitem{WHA} J. M. Wheatley, T. C. Hsu, and P. W. Anderson, Phys. Rev. B 37,
5897 (1988).
\bibitem{CSAS} S. Chakravarty, A. Sudb\o, P. W. Anderson, and
S. Strong, Science 261, 337 (1993).
\bibitem{WWS} P. B. Weigmann, Phys. Rev. Lett. 60, 821 (1988); X. G. Wen,
Phys. Rev. B 39, 7223 (1989); R. Shankar, Phys. Rev. Lett. 63, 203 (1989).
\bibitem{gauge} G. Baskaran and P. W. Anderson, Phys. Rev. B 37,
580 (1988); G. Baskaran, Phys. Scr. T27, 53 (1989).
\bibitem{dec} L. B. Ioffe and A. I. Larkin, Phys. Rev. B 39, 8988 (1989);
N. Nagaosa, Phys. Rev. Lett. 71, 4210 (1989); S. Khlebnikov, Phys. Rev. B
50, 6954 (1994); Y. B. Kim and X.-G. Wen, Phys. Rev. B 50, 8078 (1994).
In the superconducting phase, where the fermions have a gap, confinement of
charges may be restored (this question requires further study) but
in any case we expect the confinement radius to be much larger
than the coherence length of superconductivity.

\end{thebibliography}
\end{document}